\pdfoutput=1
\documentclass[11pt]{article}

\usepackage[T1]{fontenc}
\usepackage{lmodern}
\usepackage[utf8]{inputenc}
\usepackage[a4paper,margin=2.5cm]{geometry}
\usepackage{amsmath,amssymb}
\usepackage{graphicx}
\graphicspath{{./}{figures/}}
\usepackage{booktabs}
\usepackage{array}
\usepackage{caption}
\usepackage{float}
\usepackage{enumitem}
\usepackage{ragged2e}
\usepackage{authblk}
\usepackage[hidelinks]{hyperref}

\newcommand{\fitwidth}[1]{\resizebox{\ifdim\width>\linewidth\linewidth\else\width\fi}{!}{#1}}

\setlist{leftmargin=1.6em,topsep=2pt,itemsep=1pt}

\newcolumntype{R}{>{\RaggedLeft\arraybackslash}p{0.95cm}}
\title{\bfseries Bayesian Copula Directional Dependence is Cross-Network
Robust for Gene-Regulatory Pair Direction: A Benchmark Study on DREAM5}

\author[1]{Xiaoying Wei}
\author[1,$\ast$]{Clara Grazian}
\affil[1]{School of Mathematics and Statistics, University of Sydney, Sydney, NSW, Australia}
\affil[ ]{\small $\ast$\,Corresponding author: \texttt{clara.grazian@sydney.edu.au}}

\date{June 25, 2026}

\begin{document}
\maketitle
\vspace{-2.0em}
\begin{center}\small\itshape
Preprint. This note documents the empirical results accompanying our poster;
the full methodology is reported in a separate paper (in preparation).
\end{center}
\vspace{0.6em}

\begin{abstract}
\noindent
Inferring the \emph{direction} of a gene-regulatory relationship --- which gene
is the regulator and which is the target --- is harder than inferring that a
relationship exists, and most direction-inference methods are validated on a
single \emph{in silico} benchmark. We ask a different question: \emph{which method
stays reliable as the data move from a synthetic network to real organisms, and
as the sample size shrinks?} We embed a copula-based measure of directional
dependence (CDD) in a Bayesian framework that returns, for each candidate pair,
a full posterior over a directional contrast together with a credible interval,
a posterior sign-support score, and a principled no-call. We benchmark this
estimator against eight direction-inference methods --- including two established
Bayesian DAG-posterior baselines (BCDAG, BiDAG-BGe) --- on the three core DREAM5
networks (Network~1 \emph{in silico}, Network~2 \emph{S.\ aureus}, Network~3
\emph{E.\ coli}), plus an out-of-domain eukaryotic network (Network~4
\emph{S.\ cerevisiae}). Across the three core networks, Bayesian CDD is the
\textbf{only} method whose called accuracy is always $>60\%$, whose coverage is
always $>88\%$, and whose direction AUROC is always $>0.6$; every competing
method falls to chance or below on at least one network. CDD ranks first on both
real-organism networks (\emph{S.\ aureus} $87$--$90\%$, AUROC $0.94$--$0.96$;
\emph{E.\ coli} $75$--$76\%$, AUROC $0.80$--$0.81$), stays stable on the
smallest-sample network where bootstrap-interval methods collapse, and is the
only Bayesian method that is simultaneously above chance and high-coverage ---
the DAG-posterior baselines sit at chance when forced to call and make zero
confident calls under a $95\%$ posterior gate. We position CDD as a
post-screening, uncertainty-aware direction-refinement tool for candidate
regulatory pairs.

\end{abstract}

\section{Introduction}
The goal of gene-regulatory-network (GRN) inference is to recover regulatory
relationships among genes from expression data. Most benchmarks emphasise
\emph{connection} inference (does an edge exist?), but \emph{direction} inference
--- the transcription-factor~$\to$~target orientation --- is what matters for
biological interpretation and for designing perturbation experiments. Existing
direction-inference methods fall into a few families:
\begin{itemize}
\item \textbf{Network-level screeners} such as GENIE3 (random-forest feature
importance), which give accurate orientations on synthetic data but no per-pair
measure of uncertainty;
\item \textbf{Pairwise causal methods} such as ANM-GAM (residual independence),
RECI (regression-error ratio), and SLOPE/MDL, which issue a forced call for every
pair and cannot abstain;
\item \textbf{Constraint-based} methods such as the PC algorithm, which return a
partially directed graph (CPDAG) with no posterior; and
\item \textbf{Bootstrap-interval} wrappers (Bootstrap-RECI, Bootstrap-LiNGAM),
whose interval validity relies on the resampling distribution approaching the true
sampling distribution --- a large-sample property.
\end{itemize}

\noindent We see three gaps. \textbf{(i)~Uncertainty gap:} most methods provide no
per-pair posterior and cannot grade the strength of directional evidence.
\textbf{(ii)~Robustness gap (our main concern):} methods are usually validated on a
single \emph{in silico} network, so it is unknown which ones survive the move to
real organisms. \textbf{(iii)~Sample-size gap:} bootstrap intervals degrade at
small sample sizes, exactly the regime of real expression cohorts.

This note reports a head-to-head benchmark addressing all three gaps. Our estimator,
\emph{Bayesian copula directional dependence} (Bayesian CDD), turns a classical
copula directional-dependence contrast into a full posterior, supplying a credible
interval, a sign-support score, and a principled no-call. We evaluate it against
eight comparators across four DREAM5 networks. The central empirical finding is
that Bayesian CDD is the only method that is robust across all three core networks
simultaneously, and the only Bayesian method that delivers a usable, high-coverage,
above-chance, uncertainty-gated direction report. \textbf{The scope of this preprint
is deliberately empirical: it documents and makes citable the comparative results
that accompany our poster. The full estimator is specified in a separate
methodological paper (in preparation).}

\section{Method (overview)}\label{sec:method}
Copula directional dependence (CDD) compares the conditional dependence strength of
the two possible orientations of a pair $(X,Y)$ through a copula-based contrast: the
larger the dependence in one direction relative to the other, the stronger the
evidence for that orientation. Classical CDD returns this contrast as a single point
estimate with no measure of uncertainty.

We recast the contrast within a Bayesian estimation framework. For each ordered
candidate pair the procedure returns a \textbf{full posterior distribution over the
directional contrast} rather than a point estimate, from which we read off four
quantities used throughout the results:
\begin{enumerate}[label=(\roman*)]
\item a posterior point summary of the contrast;
\item a $95\%$ credible interval (equal-tailed);
\item a posterior \emph{sign-support} score --- the posterior probability that the
contrast favours one orientation over the other; and
\item a \textbf{principled no-call}: when the $95\%$ credible interval is consistent
with no directional preference, the pair is left uncalled rather than forced into a
binary label.
\end{enumerate}
We summarise performance with two paired quantities:
\[
\text{Coverage}=\frac{N_{\text{called}}}{N_{\text{total}}},\qquad
\text{Called accuracy}=\frac{N_{\text{correct}}}{N_{\text{called}}} .
\]
Coverage is the fraction of pairs on which the method is willing to commit;
called accuracy is the accuracy among those committed calls. A method can buy
accuracy by abstaining on hard pairs, so the two must always be read together.

\paragraph{Why the method section is brief.}
Because the primary paper is not yet published, the algorithmic core is intentionally
omitted here: the specific dependence functional, the regression specification, the
data-transformation step, the prior and posterior-sampling scheme, and the calibration
of the decision rule are all deferred to that paper. The present preprint should be read
as an empirical, reproducible-from-outputs benchmark report, not as a full specification
of the estimator. The one property we rely on conceptually in the results is that the
uncertainty is a \emph{model-based posterior}, obtained by exact posterior sampling under
a parametric likelihood, and therefore does not depend on large-sample asymptotics or on
the convergence of a resampling distribution --- which is what makes the small-sample
behaviour in \S\ref{sec:smallsample} possible.

\paragraph{Comparators.}
We compare against eight published methods spanning the families above: GENIE3,
ANM-GAM, RECI (point and $B{=}200$ bootstrap variants), Bootstrap-LiNGAM, SLOPE/MDL,
the PC algorithm, a sparse linear GRN (glmnet), and two Bayesian DAG-posterior
baselines, BCDAG (collapsed Gaussian-DAG MCMC) and BiDAG-BGe (order-MCMC). All are run
on the same benchmark rows; no post-hoc orientation flipping is applied to any method.

\section{Benchmark Design}\label{sec:design}
\paragraph{Networks.}
We use the four DREAM5 networks (Table~\ref{tab:nets}). Networks~1--3 (\emph{in silico},
\emph{S.\ aureus}, \emph{E.\ coli}) are the core cross-domain set; Network~4
(\emph{S.\ cerevisiae}) is the only eukaryote and is included as a deliberate
out-of-domain stress test of whether the dependence-asymmetry signal transfers across
the prokaryote/eukaryote boundary.

\begin{table}[H]\centering\small
\caption{The four DREAM5 networks.}\label{tab:nets}
\begin{tabular}{@{}lcccc@{}}
\toprule
 & Network 1 & Network 2 & Network 3 & Network 4 \\
\midrule
Organism & \emph{in silico} & \emph{S.\ aureus} & \emph{E.\ coli} & \emph{S.\ cerevisiae} \\
Genes    & 1{,}643 & 2{,}810 & 4{,}511 & 5{,}950 \\
TFs      & 195 & 99 & 334 & 333 \\
Samples  & 805 & 160 & 805 & 536 \\
Gold-standard positive edges & 4{,}012 & 518 & 2{,}066 & 3{,}940 \\
\bottomrule
\end{tabular}
\end{table}

\paragraph{Constructing direction labels.}
The DREAM5 gold standard lists only \emph{positive} edges, each recorded in the single
canonical TF~$\to$~target orientation. Used directly, every row would carry the same
label and the benchmark would be degenerate. To obtain a direction-balanced
($50\%$ correct / $50\%$ reversed) discrimination task, a portion of the edges is
re-entered in reversed orientation to supply the ``wrong-direction'' examples. We build
two complementary benchmarks per network: a \textbf{paired} design, in which every edge
appears in both orientations (a deterministic mirror construction that maximally probes
direction sensitivity), and an \textbf{independent} design, in which each edge appears
once with half reversed at random (removing any artefact the mirror pairing might
introduce). Agreement between paired and independent is itself a robustness check: a
method that genuinely discriminates direction should behave the same under both. Row
counts are 4{,}000 (N1), 500 (N2), 2{,}000 (N3) and 3{,}800 (N4) per benchmark, balanced
$50/50$ throughout.

\section{Results}

\subsection{Cross-network direction accuracy}
\begin{figure}[H]\centering
\includegraphics[width=0.85\linewidth]{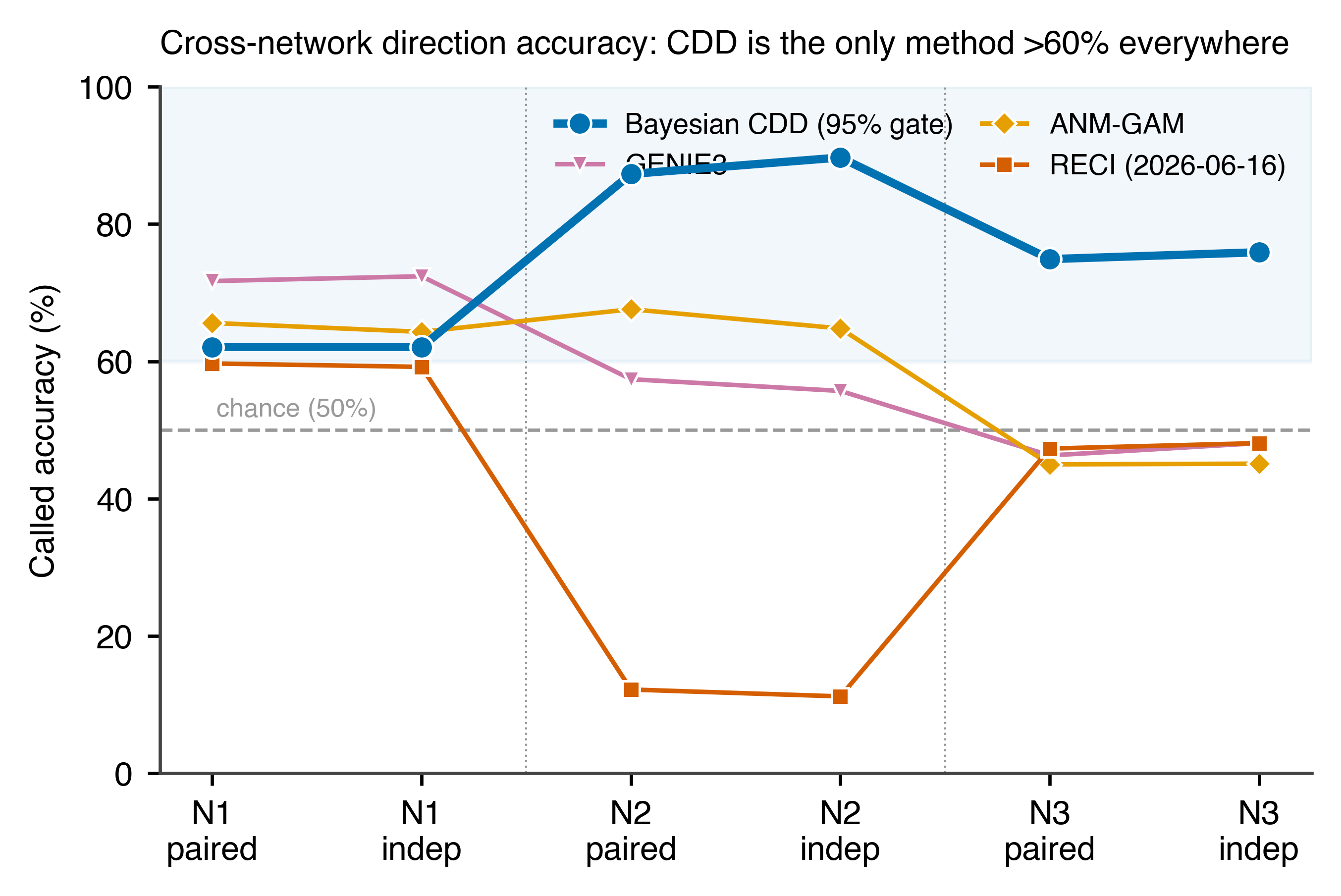}
\caption{Cross-network called accuracy (\%) across the six core network$\times$benchmark
conditions, shown as an annotated heatmap; cells are coloured diverging about the $50\%$
chance line (green above, red below). Bayesian CDD (outlined top row) is the only method
above $60\%$ in every condition, whereas competing methods each drop to chance or below on at
least one real-organism network.}
\label{fig:acc}
\end{figure}

Table~\ref{tab:acc} reports called accuracy for all methods across the four networks and
both benchmark designs; Fig.~\ref{fig:acc} visualises the three core networks. The pattern
is consistent and is the headline of this study.

\begin{table}[H]\centering\small
\caption{Called accuracy (\%): 8 methods $\times$ 4 networks (pr = paired, in = independent).
Bold marks the best per network on Networks 1--3. No post-hoc inversion is applied.
On Network~4 no cell is bolded: the nominally highest entries (PC, glmnet) are produced
only at sub-$6\%$ coverage (see text).}\label{tab:acc}
\fitwidth{
\begin{tabular}{@{}lRRRRRRRR@{}}
\toprule
Method & N1 pr & N1 in & N2 pr & N2 in & N3 pr & N3 in & N4 pr & N4 in \\
\midrule
\textbf{Bayesian CDD ($95\%$ gate)} & 62.1 & 62.1 & \textbf{87.3} & \textbf{89.7} & \textbf{74.9} & \textbf{75.9} & 50.6 & 50.3 \\
GENIE3 & \textbf{71.7} & \textbf{72.4} & 57.4 & 55.7 & 46.3 & 48.1 & 46.9 & 47.5 \\
ANM-GAM & 65.6 & 64.3 & 67.6 & 64.8 & 45.0 & 45.1 & 46.6 & 45.6 \\
RECI-poly3 & 60.0 & 58.9 & 28.8 & 26.6 & 50.9 & 50.6 & 42.5 & 43.2 \\
Bootstrap-LiNGAM ($95\%$ gate) & 59.2 & 58.7 & 6.4 & 7.0 & 64.7 & 62.7 & 58.2 & 59.2 \\
SLOPE (MDL) & 61.7 & 61.1 & 25.2 & 27.0 & 51.0 & 50.8 & 42.6 & 43.3 \\
PC algorithm & 59.0 & 58.0 & 37.5 & 50.0 & 53.3 & 52.1 & 61.1 & 55.6 \\
Sparse linear GRN (glmnet) & 54.8 & 56.4 & 41.4 & 42.0 & 42.1 & 40.5 & 56.6 & 51.8 \\
\bottomrule
\end{tabular}}
\end{table}

\noindent\textbf{Network 1 (\emph{in silico}).} GENIE3 is highest ($71.7$/$72.4$); CDD is
mid-pack ($62.1$). \textbf{Network 2 (\emph{S.\ aureus}).} CDD ranks first ($87.3$/$89.7$),
leading the next-best method (ANM-GAM) by more than $20$ percentage points, while RECI sits
near $27\%$ and Bootstrap-LiNGAM near $7\%$. \textbf{Network 3 (\emph{E.\ coli}).} CDD ranks
first again ($74.9$/$75.9$), while GENIE3 and ANM-GAM fall \emph{below} chance.
\textbf{Across Networks 1--3, CDD is the only method that stays above $60\%$ everywhere}
(Fig.~\ref{fig:acc}). Methods that excel on the synthetic network do not transfer: choosing a
method from Network~1 alone and applying it to \emph{S.\ aureus} would be badly misleading.

\noindent\textbf{Network 4 (eukaryotic boundary).} Behaviour on Network~4 is qualitatively
different: every full-coverage method collapses toward chance, including CDD
($50.6$/$50.3$). The two nominal ``leaders'' (PC $61.1$, glmnet $56.6$) reach those numbers
only by abstaining on almost every pair (sub-$6\%$ coverage). We report Network~4 honestly as
an out-of-domain limit: yeast transcriptional regulation is more layered and combinatorial
than the prokaryotic regulation of N2/N3, so this likely reflects a genuine domain shift
rather than a mere instability. Our cross-network robustness claim is therefore scoped to
the \emph{in silico} and prokaryotic networks (N1--N3).

\subsection{Coverage and uncertainty-aware comparison}
\begin{figure}[H]\centering
\includegraphics[width=0.85\linewidth]{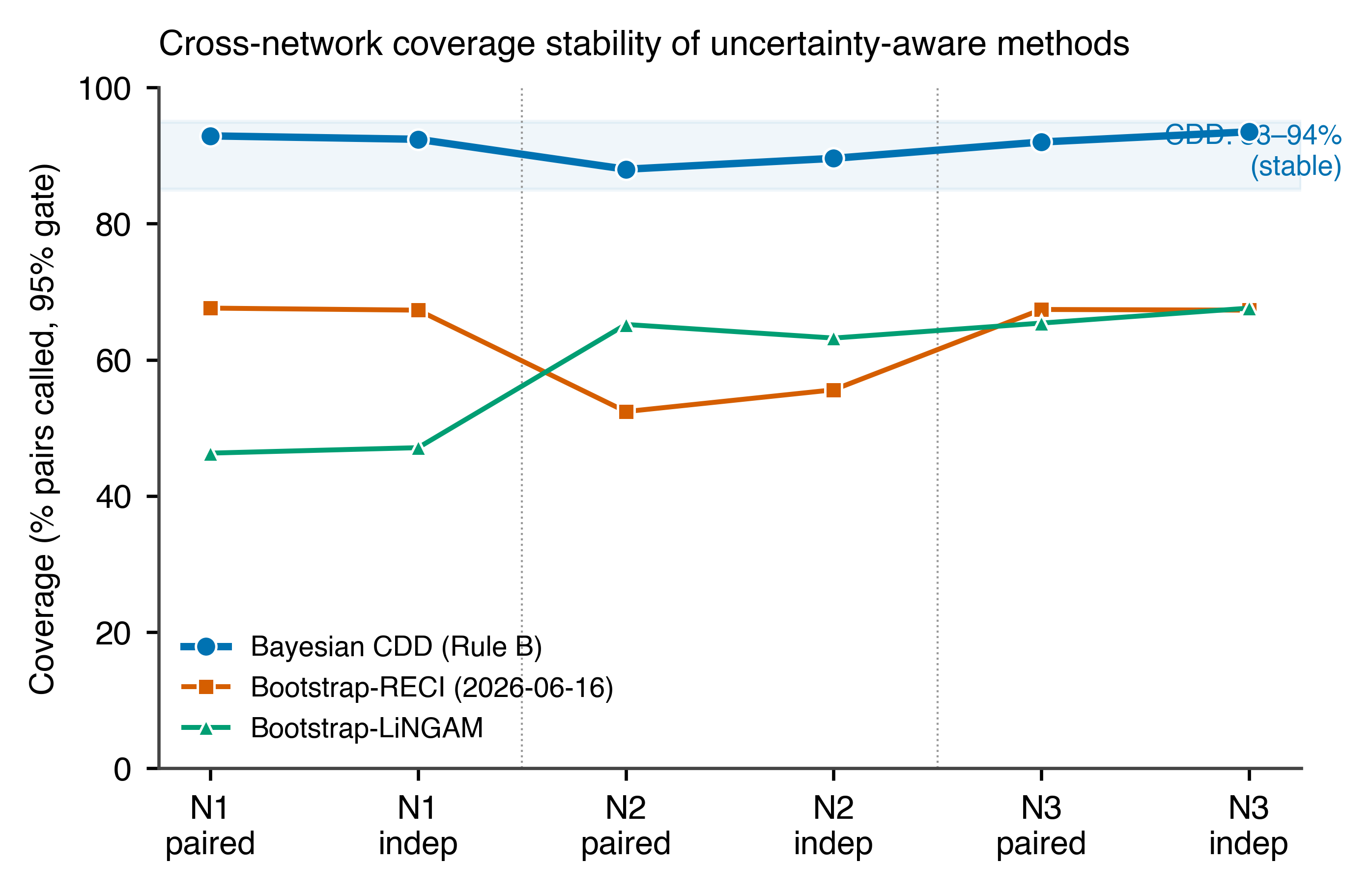}
\caption{Cross-network coverage (\%) of the interval-capable methods, shown as an annotated
heatmap (darker $=$ higher coverage). Bayesian CDD (outlined top row) holds a uniformly high
$88$--$94\%$ across every condition, whereas the bootstrap-interval methods are lighter and
uneven, swinging sharply across networks.}
\label{fig:cov}
\end{figure}

The accuracy numbers only become interpretable alongside coverage. Table~\ref{tab:cov}
compares the three methods that produce an interval-based no-call under a common $95\%$
gate, and Fig.~\ref{fig:cov} plots coverage stability.

\begin{table}[H]\centering\small
\caption{Accuracy and coverage of the interval-capable methods under a $95\%$ gate
(RECI rows are the re-run reported in \S\ref{sec:notes}).}\label{tab:cov}
\fitwidth{
\begin{tabular}{@{}llRRRRRR@{}}
\toprule
Method & metric & N1 pr & N1 in & N2 pr & N2 in & N3 pr & N3 in \\
\midrule
\textbf{Bayesian CDD} & accuracy & 62.1 & 62.1 & 87.3 & 89.7 & 74.9 & 75.9 \\
\textbf{Bayesian CDD} & coverage & 92.9 & 92.4 & 88.0 & 89.6 & 92.0 & 93.5 \\
Bootstrap-RECI & accuracy & 59.7 & 59.2 & 12.2 & 11.2 & 47.3 & 48.1 \\
Bootstrap-RECI & coverage & 67.6 & 67.3 & 52.4 & 55.6 & 67.4 & 67.3 \\
Bootstrap-LiNGAM & accuracy & 59.2 & 58.7 & 6.4 & 7.0 & 64.7 & 62.7 \\
Bootstrap-LiNGAM & coverage & 46.3 & 47.1 & 65.2 & 63.2 & 65.4 & 67.6 \\
\bottomrule
\end{tabular}}
\end{table}

\noindent CDD holds coverage in a tight $88$--$94\%$ band (range $<6$ pp) across every
network, whereas Bootstrap-RECI ($52$--$68\%$) and Bootstrap-LiNGAM ($46$--$68\%$)
fluctuate by $20$ percentage points or more. CDD therefore commits to a direction on the
large majority of pairs \emph{and} is right more often when it does, on every core network.

\subsection{Small-sample stability}\label{sec:smallsample}
Network~2 (\emph{S.\ aureus}) is the smallest dataset --- 500 pairs and only 160 expression
samples per gene --- and is the decisive test of sample-size robustness. There CDD attains
$87$--$90\%$ accuracy at $\sim 88\%$ coverage (still the best of all methods), while
Bootstrap-RECI coverage collapses to $52$--$56\%$ at $11$--$12\%$ accuracy and
Bootstrap-LiNGAM accuracy falls to $6$--$7\%$. This is the behaviour predicted by the
nature of the two kinds of uncertainty: bootstrap intervals rely on the resampling
distribution approaching the true sampling distribution, which fails at small samples,
whereas CDD's model-based posterior remains calibrated at any sample size --- a small
sample merely widens the credible interval and converts borderline pairs into no-calls,
rather than producing confident wrong calls. In a controlled simulation with a known
directional asymmetry (100 replicates per sample size), the same mechanism is visible:
the $95\%$-credible-interval power is $100\%$ at $N\ge500$ and $0\%$ at $N=100$ --- i.e.\
the method \emph{abstains} when the data are insufficient instead of guessing.

\subsection{Direction AUROC and AUPR}
\begin{figure}[H]\centering
\includegraphics[width=0.92\linewidth]{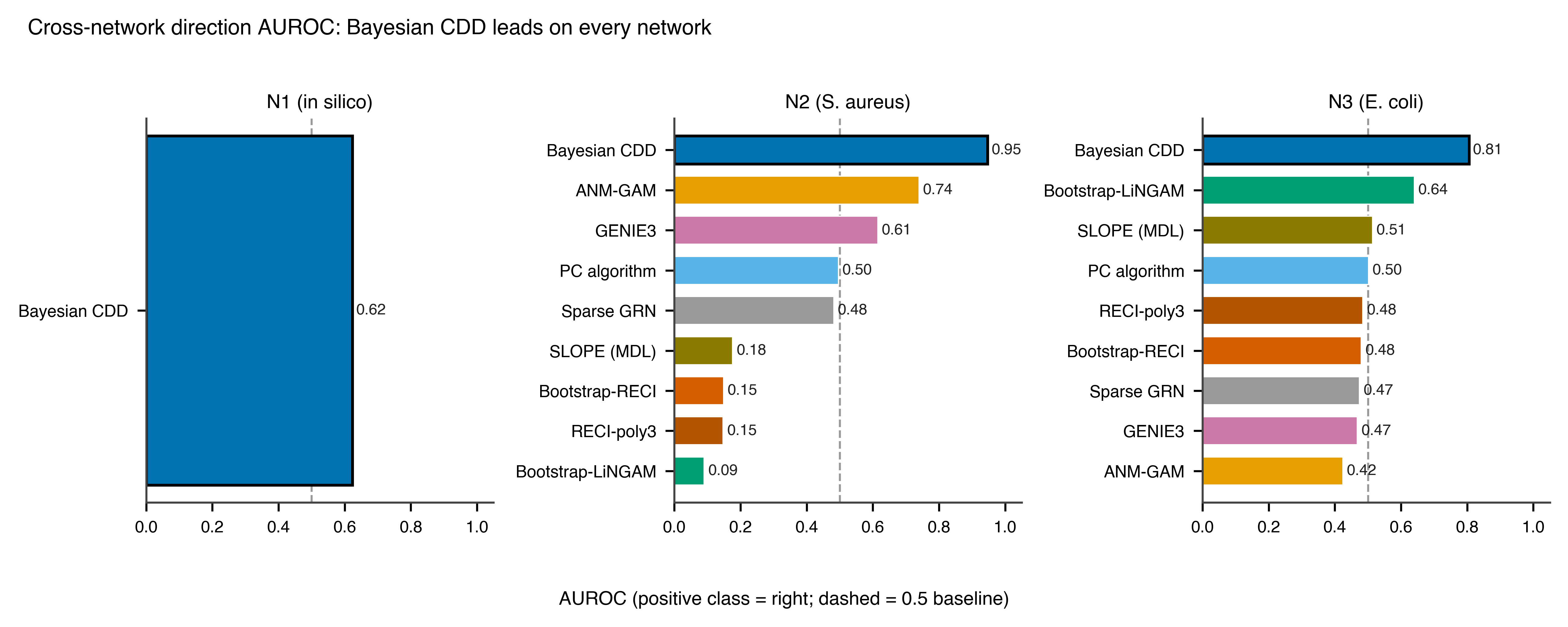}
\caption{Cross-network direction AUROC (per-network bars; paired and independent averaged;
dashed line $=0.5$ no-signal baseline). Bayesian CDD is the top bar on every network.}
\label{fig:auroc}
\end{figure}

Coverage and called accuracy depend on a decision threshold; AUROC and AUPR remove that
dependence by scoring the full ranking. Treating ``correct direction'' as the positive
class (a balanced $50/50$ task, so the AUPR baseline is $0.50$) and using each method's
native directional score, CDD is the only method with AUROC $>0.6$ on all three core
networks: N1 $0.62$, N2 $0.94$--$0.96$, N3 $0.80$--$0.81$ (Fig.~\ref{fig:auroc}). Several
comparators fall \emph{below} $0.5$ on a real-organism network (e.g.\ on N2, RECI $\approx0.14$
and LiNGAM $\approx0.09$; on N3, ANM-GAM $\approx0.42$ and GENIE3 $\approx0.46$), meaning their
native ranking is anti-aligned with the true direction there. These values are reported as-is,
with no post-hoc flipping.

\subsection{Comparison against Bayesian DAG baselines}\label{sec:bayes}
\begin{figure}[H]\centering
\includegraphics[width=0.92\linewidth]{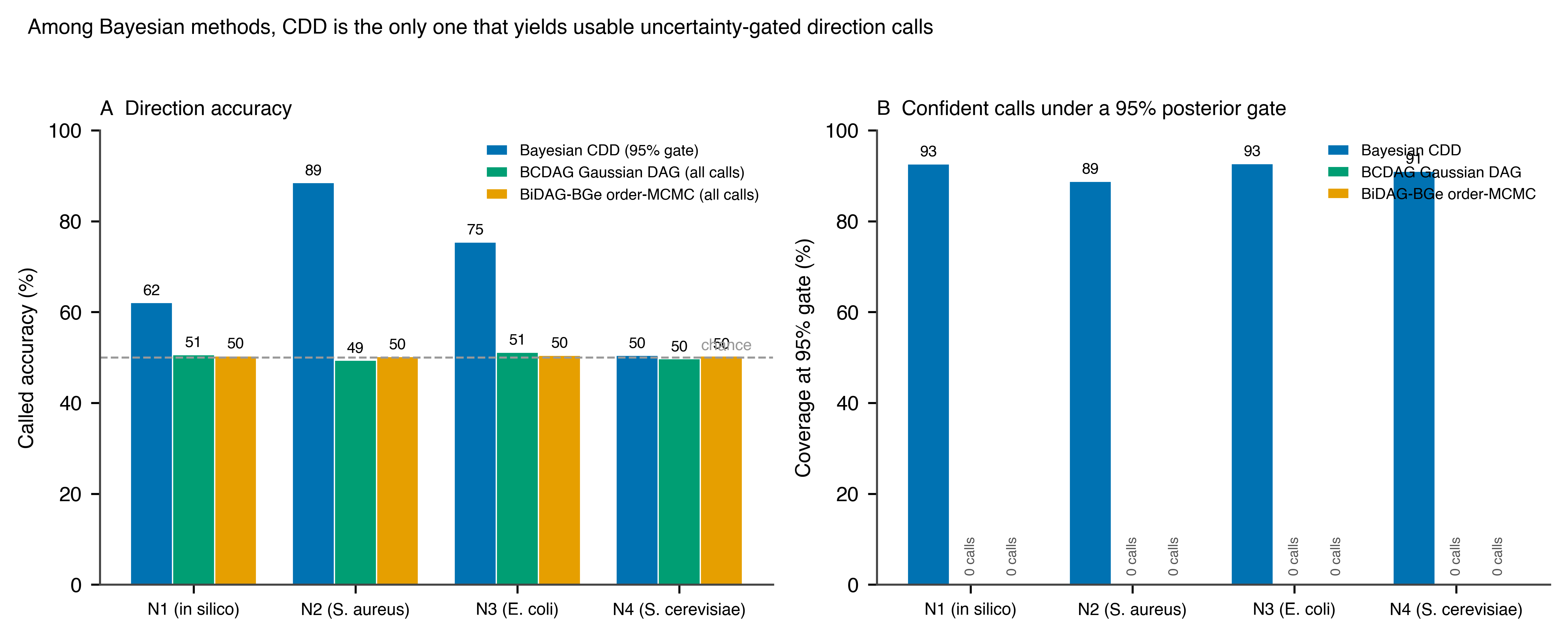}
\caption{Bayesian-method comparison. \textbf{(A)} Called accuracy: BCDAG and BiDAG sit at
chance even when forced to call every pair; CDD is well above chance on N1--N3.
\textbf{(B)} Coverage under the native $95\%$ posterior gate: the DAG baselines produce zero
confident calls everywhere, while CDD keeps $88$--$93\%$ coverage. Values average the paired
and independent benchmarks.}
\label{fig:bayes}
\end{figure}

Because our central claim is a \emph{Bayesian} one --- a full posterior with a principled
no-call --- the most demanding test is against other Bayesian methods that also produce a
posterior over edge orientation. We ran two established Bayesian structure-learning baselines
on identical benchmark rows: BCDAG (collapsed Gaussian-DAG MCMC) and BiDAG-BGe (order-MCMC).
Both yield a posterior edge-inclusion (hence orientation) probability per pair --- the natural
Bayesian analogue of CDD's sign-support score. We compare on equal terms: forcing a call on
every pair, and applying the native $95\%$ posterior gate (Table~\ref{tab:dag},
Fig.~\ref{fig:bayes}).

\begin{table}[H]\centering\small
\caption{Bayesian head-to-head across the four networks (paired and independent averaged).
Coverage is the fraction of pairs called under the native $95\%$ posterior gate.}\label{tab:dag}
\fitwidth{
\begin{tabular}{@{}llRRRR@{}}
\toprule
Bayesian method & metric & N1 & N2 & N3 & N4 \\
\midrule
\textbf{Bayesian CDD ($95\%$ gate)} & called accuracy & \textbf{62.1} & \textbf{88.5} & \textbf{75.4} & 50.5 \\
\textbf{Bayesian CDD ($95\%$ gate)} & $95\%$ coverage & \textbf{92.6} & \textbf{88.8} & \textbf{92.7} & \textbf{91.0} \\
\addlinespace
BCDAG Gaussian DAG & accuracy (all calls) & 50.7 & 49.4 & 51.2 & 49.7 \\
BCDAG Gaussian DAG & $95\%$ coverage & 0 & 0 & 0 & 0 \\
\addlinespace
BiDAG-BGe order-MCMC & accuracy (all calls) & 50.4 & 50.2 & 50.5 & 50.3 \\
BiDAG-BGe order-MCMC & $95\%$ coverage & 0 & 0 & 0 & 0 \\
\bottomrule
\end{tabular}}
\end{table}

\noindent The contrast is stark on two fronts. \textbf{(i)~Accuracy:} forced to call every
pair, both DAG-posterior baselines sit at chance on every network ($49$--$53\%$) --- their
global edge posteriors carry essentially no pair-level direction signal here --- while CDD is
far above chance on N1--N3. \textbf{(ii)~Usable uncertainty:} under their own $95\%$ posterior
gate, BCDAG and BiDAG make \emph{zero} confident calls on all four networks (coverage $=0$),
because the edge-level credible region almost never excludes the null orientation; CDD instead
retains $88$--$93\%$ coverage at the same nominal confidence. Among Bayesian methods, CDD is
therefore the only one that delivers a usable, high-coverage, above-chance, uncertainty-gated
direction report. The mechanism is intuitive: a whole-DAG posterior must hedge over global
graph uncertainty, whereas CDD conditions on an already-established pair and asks only the
local orientation question, concentrating posterior mass where the direction signal lives.

\section{Discussion}
\paragraph{What the benchmark shows.}
Five findings stand out. \textbf{(1)}~Bayesian CDD is the only method robust across all three
core networks at once, and ranks first on both real-organism networks. \textbf{(2)}~A no-call
is a result, not a failure: CDD's abstention rate is a stable $6.6$--$12.0\%$ across species and
data scales, so coverage stays high while the report remains graded. \textbf{(3)}~CDD stays
reliable at small samples because its uncertainty is a model-based posterior rather than a
resampling interval. \textbf{(4)}~CDD is best understood as a \emph{second-stage} method: a
screener such as GENIE3 nominates candidate edges, and CDD then supplies posterior direction
evidence on them. \textbf{(5)}~Single-network evaluation is misleading --- the method that wins
on the synthetic network (GENIE3) drops below chance on \emph{E.\ coli}.

\paragraph{Limitations.}
The decision rule is calibrated on the benchmark rather than validated against ground-truth
perturbation (e.g.\ CRISPR) data. Pairwise evaluation treats each pair in isolation and cannot
account for indirect effects, confounding, or feedback. Network~2's small scale gives limited
statistical power --- the same regime in which the small-sample-stability argument is made, so
it should be read as suggestive rather than definitive. CDD is a direction-refinement tool, not
an edge-discovery tool: as a connection detector its AUC is below GENIE3 and a sparse linear
GRN, and it should only be applied to pairs whose regulatory status is already established.
Finally, on the eukaryotic Network~4 CDD sits at chance and the gate does not help; the
robustness claim does not yet extend to eukaryotes, and that generalisation is a priority for
future validation.

\paragraph{Outlook.}
Natural extensions include dynamic (time-series) and network-conditioned variants to mitigate
confounding, faster posterior approximation, informative priors derived from screener
importances, validation of the decision rule on perturbation data, and deeper eukaryotic
evaluation suited to combinatorial regulation.

\section{Conclusion}
On the DREAM5 direction benchmark, Bayesian copula directional dependence is the only method
that is simultaneously cross-network robust (called accuracy $>60\%$, coverage $>88\%$, AUROC
$>0.6$ on all three core networks), first-ranked on both real-organism networks, stable in the
small-sample regime where bootstrap-interval methods collapse, and --- among Bayesian methods
--- the only one that is at once above chance and high-coverage. We position it as a
post-screening, uncertainty-aware direction-refinement tool for candidate regulatory pairs.
The full estimator and its theoretical justification are reported in a separate paper
(in preparation).

\section*{Data and reproducibility notes}\label{sec:notes}
\addcontentsline{toc}{section}{Data and reproducibility notes}
\small
All results are computed on the public DREAM5 expression data and gold standards. The RECI /
Bootstrap-RECI rows were re-run on the fixed benchmark rows; only the RECI family was updated in
this revision, the other comparators are retained from the original formal benchmark, and
Bootstrap-LiNGAM was not re-run. AUROC/AUPR use each method's native directional score with the
positive class defined as the correct orientation and no post-hoc flipping. The Bayesian
DAG-posterior baselines (BCDAG, BiDAG-BGe) were run on exactly the same benchmark rows as the
cross-network direction benchmark. Numerical values in all tables are reproducible from the
per-prediction output files underlying this study.

\section*{Acknowledgements}
\small
We thank colleagues at the School of Mathematics and Statistics, University of Sydney, for
helpful discussions.


\end{document}